\documentclass[12pt,fleqn]{article}
\usepackage{times}
\usepackage{graphics,latexsym}
\usepackage{amsmath}
\usepackage{amssymb}
\usepackage[pdftex]{graphicx}
\usepackage{epstopdf}
\usepackage{lscape}
\usepackage{rotating}
\usepackage{color}

\newcommand{\ba}{\begin{array}}
\newcommand{\ea}{\end{array}}

\begin{document}
\newcommand{\be}{\begin{equation}}
\newcommand{\ee}{\end{equation}}
\newcommand{\bc}{\begin{center}}
\newcommand{\ec}{\end{center}}
\newcommand{\bdm}{\begin{displaymath}}
\newcommand{\edm}{\end{displaymath}}
\newcommand{\ds}{\displaystyle}
\newcommand{\p}{\partial}
\newcommand{\INT}{\int\limits}
\newcommand{\SUM}{\sum\limits}
\newcommand{\bfm}[1]{\mbox{\boldmath $ #1 $}}
\renewcommand{\theequation}{\arabic{section}.\arabic{equation}}

\title{ \bf
Iontophoretic transdermal drug  delivery:  \\
a multi-layered approach
}

%\author{Third Author}{author3@example.org}{2}

%\affiliation{number}{department}{university}

\author{
{\em  Giuseppe Pontrelli$^{a}$, Marco Lauricella$^{a}$, Jos\'e A. Ferreira$^{b}$, Gon\c{c}alo Pena$^{b}$, } 
\footnote{Corresponding author, E-mail: {\tt giuseppe.pontrelli@gmail.com}}
\vspace{5mm}\\
$^{b}$Istituto per le Applicazioni del Calcolo - CNR \\
Via dei Taurini 19 -- 00185 Rome, Italy \\
\vspace{1mm}\\
$^{a}$CMUC, Department of  Mathematics\\
University of Coimbra, Portugal 
}

\maketitle

\begin{abstract} 
We present a multi-layer mathematical model to describe the the transdermal 
drug release
from an iontophoretic system.  
The  Nernst-Planck equation  
describe the basic convection-diffusion process, 
 with  the electric potential
obtained by Laplace equation. These equations
are  complemented with suitable interface and boundary conditions in a 
multi-domain. The stability of the
mathematical problem is discussed in different scenarios and a 
finite-difference 
method is used to solve the coupled system. Numerical
experiments are included to illustrate the drug dynamics under different
conditions.
\end{abstract}

\vspace*{2ex}\noindent\textit{\bf Keywords}: drug release, iontophoresis, Nernst-Plank equations, finite-difference methods.

\section{Introduction}

Traditional transdermal drug delivery  (TDD) systems are based on the transport of
therapeutic agents across the skin by passive diffusion.  Despite being the subject of intense research over the past years, it is still unclear  the exact mechanism of  release and it is often difficult to accurately predict the drug kinetics \cite{Prautnitz, Trommer}. Toxicity can arise if an excessive amount of drug is delivered, or if it is released too quickly.  On the other hand, if the drug is delivered at slow rate, or at sufficiently
low concentrations, the
therapeutic effect vanishes \cite{Prausnitz_generations}. The success of the TDD is therefore dependent
on the amount of drug, rate of release and binding to cell receptors.   However, the skin has unique structural
and physico-chemical properties which differentiate it from other
bio-membranes \cite{mill}. The solutes which can be delivered 
by transdermal route are limited due to the
excellent barrier properties of the stratum corneum, the outermost layer of the
epidermis. As a matter of fact, most drugs
delivered by the conventional systems  are constituted by small and highly lipophilic molecules \cite{Pang}. To increase  skin's drug transport  and overcome the barrier properties of the stratum corneum, innovative technologies have been developed, based on the use of drug transport enhancers. Two main classes of these have been proposed in the literature. The first one is composed by chemical penetration enhancers including surfactants, fatty acids, esters and solvents.  These products can have side effects and toxicological implications and appear to be restricted at the present time to experimental strategies. The second group of enhancers, the physical ones, apply an external energy to raise the drug delivery into and across the skin or change the structure of the skin itself and perforate the stratum corneum creating microchannels to facilitate the drug transport  \cite{Prausnitz_generations,Perumal}.

Drug delivery devices   based on the application of an external energy source includes electrically assisted systems where the applied potential generates an additional driving force for the drug motion (see \cite{pra,becker2} and references therein).
Among such systems, here we are interested in iontophoretic TDD that several studies  indicate as effective in topical and systemic
administration of drugs. It can be used to treat, for instance, dermal analgesia, management of migraine \cite{Ita_Drug_Targeting},  or  acute postoperative pain \cite{Power_Anaesthesia}. More recent applications of iontophoresis have been considered, for instance, in cancer treatment \cite{Byrne,Komuro}.  
 In iontophoretic TDD, a charged drug is initially dispersed in a reservoir (or vehicle) which is in contact with the skin, the target tissue. The electric field is generated by a potential of low intensity ($\approx 1V$) and applied over a limited period of time to
prevent any skin damage.  
The iontophoretic transport of an active agent across the skin can be expressed in terms of three independent contributions: 
passive diffusion due to a
chemical gradient, electromigration due to an electric potential
gradient and, with a minor effect, solute kinetics due to convective 
solvent flow (electroosmosis)\cite{Gratieri,Pignatello}. Moreover, 
an increased skin permeability arises from changes
in the structure of the skin caused by current flow  \cite{pli}. 
 It should be remarked that chemical reactions are also present in TDD, such
as  drug degradation,  binding and unbinding due to the drug affinity with the polymer chains of the reservoir and/or the target tissue. In this work we are mainly interested in the physical transport and the other effects are 
neglected.

Mathematical modelling of iontophoretic process allows to predict drug  release from the vehicle and its transport into the target tissue and offers insights
into the factors governing drug delivery, such as  the duration of applications and their frequency \cite{kal}. 
 For traditional TDD, the coupling between the diffusion process in the reservoir and in the target tissue has been considered in  \cite{silvia_coupling,pon,Simon_Loney}. In the
majority of TDD models for iontophoretic systems,  a constant
flux enters  the target tissue -- composed of one layer only -- and the role of the reservoir of finite capacity is neglected \cite{Perumal,Gratieri}. In \cite{Jaskari},  the authors consider a two-compartment diffusion model to describe the passive drug evolution while  a one layer model is used when the electric field is applied.  Tojo  has proposed a more general model for iontophoresis incorporating time-dependence, drug binding and metabolism as well as the convective
flow term described above \cite{Tojo}. Pikal developed a
relationship between flux enhancement by treating the process as a simple mass
transfer through aqueous channels \cite{Pikal}. A strategy to combine mathematical modelling  with in-vivo and in-vitro data has been recently proposed in \cite{Simon_ita}. However, none of the above models considers the composite structure of the skin: this aspect has a crucial importance since the drug transport critically relates to the local diffusive properties and, even more importantly, the potential field relies  on the layer-dependent electrical conductivities. 
In this paper we overcome this drawback by considering a coupled diffusion model to describe the drug release from a vehicle in a multi-layered dermal tissue under the action of an electric field.  
It is well accepted that the skin has a inhomogenous structure, 
being  composed of several layers with different thickness 
and physico-chemical-electrical properties.  The drug transport in this composite medium is described by Fick's law for the passive diffusion and by the Nernst-Planck equation for the convective transport induced by the potential gradient. It results a number of coupled convection-diffusion equations  defined in a multi-domain, complemented by suitable conditions on the contact surface and on the interfaces between the tissue layers.

 The paper is organized as follows. In Section \ref{sec::multi_layer} we introduce our vehicle-skin  multi-layer physical domain in a general framework. 
The coupled iontophoretic model is defined in Section \ref{sec::transdermal_ionto}. The mathematical problem consists of convection-diffusion equations for the drug concentration  coupled with  the Laplace equation for the electric potential. Due to the material contrast of the layers, we end up with a stiff mathematical problem.  The complex system is analyzed in Section \ref{sec::analysis} where two qualitative results concerning its stability are presented. After a suitable  nondimensionalization of the initial boundary value problem in Section \ref{sec::nondimensional}, a finite-difference scheme is described in Section \ref{sec::FD_scheme} and finally in section \ref{sec::numerical_results} some numerical experiments of TDD are presented and discussed.

\section{A multi-layer model for the coupled vehicle-skin system}\label{sec::multi_layer}
\setcounter{equation}{0}
Let us consider a TDD system constituted by {\em i)} a thin layer containing a drug ({\em the vehicle})\footnote{This can be 
the polymeric matrix of a transdermal patch, or a gel film, or an ointment rub on the skin surface, and acts as the drug 
reservoir.} and {\em  ii)} the skin where the drug is directed to, separated by a protecting film or a semi-permeable membrane.
This constitutes a coupled system with an imperfect contact interface for the mass flux
 between the two media.  Because most of the
mass dynamics occurs along the direction normal to the
skin surface, we restrict our study to
a simplified 1D model. In particular, we consider
a line crossing the vehicle and the skin, pointing inwards, and
  a Cartesian coordinate $x$ is used along it.

It is well recognized that the skin has a typical composite structure, constituted by a sequence of
contiguous layers of different physical properties and thickness, with drug capillary clearance (washout)
 taking place  at the end of it
(see %\cite{guy}
 \cite{mill} for
an anatomic and physiological description) (figs 1 - 2).
The vehicle (of thickness $\ell_0$) and the  skin layers (of thicknesses 
$\ell_1, \ell_1,... ,\ell_n $)  are treated as macroscopically homogeneous porous media.
Without loss of generality, let us assume that $x_0=0$
is the vehicle-skin interface.
In a general 1D framework, let us consider a set of intervals
$[x_{i-1},x_i], \, i=0,1,\dots,n$, having thickness
$\ell_i=x_i-x_{i-1}$, modelling the vehicle (layer 0) and the skin (layers $1,2,\dots,n$), with $L=\SUM_i \ell_i$ the overall skin thickness ($x_n=L$)
(fig. 3).

     At the initial time ($t=0$), the drug is contained only in the
vehicle, distributed with maximum
 concentration $C$
and, subsequently, released into the skin.  Here, and throughout
 this paper, a mass
 volume-averaged concentration $c_i(x,t)$ ($mg/cm^3$) is considered in
each layer $\ell_i$.
Since the vehicle
is made impermeable with a backing protection, no mass flux pass through
the boundary surface at $x=x_{-1}=-\ell_0$. 
 Strictly speaking, in a diffusion dominated problem the concentration vanishes
 asymptotically at infinite distance. However, for computational purposes,
the concentration is damped out (within a given tolerance)
over a finite distance at a given time. Such a length, (sometimes termed as {\em penetration distance}), critically depends on the diffusive properties of the layered medium \cite{pon2}. At the right end $x=L$ all drug is assumed washed out from capillaries and a sink condition ($c_n=0$)
is imposed.

\section{Transdermal iontophoresis}\label{sec::transdermal_ionto}
\setcounter{equation}{0}
To promote TDD, an electric field is locally applied 
in the area where the therapeutic agent has to be released ({\em 
iontophoresis}):
one electrode is placed in contact with the vehicle and the other is in contact with the target tissue. To simplify we assume that the drug is positively charged, the anode  is at $x=x_{-1}$ and the cathode   is at $x=x_n$ (fig. 3). 
Let $\Psi_0$ and $\Psi_1$, $\Psi_0 > \Psi_1$,  be the correspondent applied potential at the endpoints.
%and let $\Delta \Psi=\Psi_0-\Psi_1$  the applied potential in the
%multi-layer system. \par
By mass conservation, the concentration 
satisfies the following equation:
\be
{\p c_i \over \p t} + \nabla \cdot J_i=0  \label{er4}
\ee
and in each layer $i$  the mass flux is defined by the
 Nernst-Planck flux equation \cite{Gratieri}:
\be
J_i= - D_i  \nabla c_i  -u_i c_i \nabla \phi_i,    \,  i=0,1,\dots,n \label{es1}
\ee
%with $\epsilon$ the combined effective porosity and tortuosity
%($\approx 0.15$ (Li,2004)
where $\phi_i$ is the electric potential in layer $i$ and 
the convective (electroosmotic) term is omitted. Eqn. (\ref{es1}) is
the generalized Fick's first law with an additional driving 
force which is proportional to
the electric field.
The electric mobility is related to the diffusivity coefficient through
the Einstein relation:
\be
u_i= {D_i z_i F \over R T}, \,  i=0,1,\dots,n \label{es2}
\ee
($cm^2 \, V^{ -1}\, s^{-1}$), where $z_i$ ion valence, $F$ the Faraday constant, $R$ the gas constant, $T$ the
absolute temperature.
The boundary conditions are
\begin{align}
&J_0=0        \qquad \mbox{at} \quad  x=-\ell_0=-x_{-1} \qquad \mbox{(impermeable backing)}  \label{ety1}  \\
&c_n= 0       \qquad \mbox{at} \quad  x=x_n  \qquad \mbox{(sink condition due to the capillary washout)}  \label{ety2}
\end{align}
The last condition arises because, in deep skin, drug is uptaken by capillary
network and is lost in the systemic circulation: we refer to this as systemically absorbed (shortly ``absorbed'' ) drug.
At $x=0$ we  impose the matching of the total fluxes and that they are proportional to the jump of concentration (Kedem-Katchalski eqn):
\be
J_0 = J_1 =  P(c_0 - c_1)  \qquad  \mbox{at} \quad x=0   \label{et1}
\ee
with $P$(cm/s) a mass transfer coefficient
(includes a drug partitioning and a mass flux resistance).
At the other interfaces we assume a perfect contact and continuity of concentrations
and fluxes:
\be
J_i =J_{i+1}  \qquad \qquad
c_i =c_{i+1}  \qquad  \mbox{at} \quad x=x_i , \qquad\quad i=1,2,\dots,n-1   \label{ex2}
\ee
The initial conditions are set as:
\be
c_0(x,0)=C, \qquad c_i(x,0)=0 \qquad  i=1,2,\dots,n \label{etu}
\ee

\subsection{The electric potential field} 
To solve equations (\ref{er4})--(\ref{es2}), in some cases the potential $\phi$
 is assigned, but  in this multi-layer system
we find it as
solution of the Poisson equation:
\begin{align}
&\nabla \cdot (\sigma_i \nabla \phi_i)= \sum_k z_k c_k   \simeq 0 \qquad  i=0,1,\dots,n  \label{er1} \\
&  \phi_0=\Psi_0    \qquad  \mbox{at} \quad x=-x_{-1}  \label{er2} \\
& \phi_n=\Psi_1      \qquad  \mbox{at} \quad x=x_n  \label{er3}
\end{align}
with $\sigma_i ( \ds{cm^{-1} \, \Omega^{-1}}) $ the electrical conductivities
 in the layer $i$  \cite{becker2}. 
At the interfaces we assume an electrically perfect contact and we have continuity of potential
and fluxes:
\be
-\sigma_i \nabla \phi_i  = -\sigma_{i+1}  \nabla \phi_{i+1}  \qquad \qquad
\phi_i =\phi_{i+1}  \qquad  \mbox{at} \quad x=x_i \qquad i=0,1,\dots,n-1   \label{et4}
\ee
%Note that
%the concentration is depending on the electric potential, but this is independent (for a dilute solution) of the concentration.
It is straightforward to verify that
 the exact solution of  the problem (\ref{er1})--(\ref{et4}) is
\be
 \phi_i(x)=a_i x + b_i    \qquad i=0,1,\dots,n  \label{eyu}
\ee
with the expressions of $a_i (V/cm)$ and $b_i (V)$ are computed in the specific
cases (see sect. 7).

%(units of $a_i$ are V/cm,  of $b_i$ are V, so that $ \phi$ has the unit of volt)

\section{Energy estimates}\label{sec::analysis}
\setcounter{equation}{0}
In this section we establish some energy  estimates in the two cases of finite $P>0$ (subsection \ref{subsec::imperfect}) and in the limit case $P \rightarrow \infty$  (subsection \ref{subsec::perfect}) at interface condition (\ref{et1}), that leads
to upper bound of total drug masses. These results will be used
to obtain stability and uniqueness results under a condition on the applied potential and on the diffusivities.
We point out that our analysis concerns the global energy as well as the total mass of drug in the whole system, and cannot be carried out in the single layers.

\subsection{ The general case: imperfect contact}\label{subsec::imperfect}
Let $(\cdot,\cdot)_i$ be the usual inner product in $L^2(x_{i-1},x_i)$ and $\|\cdot\|_i$ the corresponding norm, $i=0,\dots,n.$ From (\ref{er4}) we deduce
\begin{equation}\label{stab1}
\sum_{i=0}^n \left(\frac{\partial c_i}{\partial t},c_i \right)_i
+\sum_{i=0}^n(\nabla \cdot J_i,c_i)_i=0.
\end{equation}
Combining (\ref{stab1})  with the boundary conditions (\ref{ety1})--(\ref{ety2}) and the interface conditions (\ref{et1}), (\ref{ex2}) we get
\begin{equation}\label{stab2}
\frac{1}{2}\xi'(t)+\sum_{i=0}^n D_i\|\nabla c_i\|_i^2+\sum_{i=0}^n(v_ic_i,\nabla c_i)_i
= -P\big(c_0(0,t)-c_1(0,t)\big)^2
\end{equation}
where $v_i=u_i\nabla \phi_i,$  is the convective velocity %(cm/s) 
induced by the potential field, and
 $ \xi(t)=\SUM_{i=0}^n\|c_i(t)\|_i^2$ is the total energy.  By the Young inequality we have:
$$(v_ic_i,\nabla c_i)_i\geq -\frac{1}{4\epsilon^2}v_i^2\|c_i\|_i^2-\epsilon_i^2\|\nabla c_i\|_i^2,$$
for any $\epsilon_i\neq 0.$ Then from (\ref{stab2}) we get
\begin{equation}\label{stab3}
\frac{1}{2}\xi'(t) +\sum_{i=0}^n \Big(D_i-\epsilon_i^2 \Big)\|\nabla c_i\|_i^2-\sum_{i=0}^n\frac{v_i^2}{\epsilon_i^2}\|c_i\|_i^2
\leq  -P\big(c_0(0,t)-c_1(0,t)\big)^2
\end{equation}
Taking $\epsilon_i^2=D_i$ in (\ref{stab3}) we obtain:
\begin{equation}\label{stab4}
 \xi'(t)  -2 \sum_{i=0}^n\frac{v_i^2}{D_i}\| c_i\|_i^2
\leq  -2P\big(c_0(0,t)-c_1(0,t)\big)^2.
\end{equation}
Finally,  (\ref{stab4}) leads to
\begin{equation}\label{stab5}
\xi(t)\leq e^{{\alpha} t}\Big(\xi(0)-2P\int_0^t
 e^{-{\alpha} s}
 \big(c_0(0,s)-c_1(0,s)\big)^2ds\Big).
\end{equation}
with  $  \alpha= 2 \max\limits_{i=0,\dots,n}\ds\frac{v_i^2}{D_i}.$
The upper bound (\ref{stab5}) shows that the initial boundary value problem (\ref{er4})-- (\ref{etu})
is stable for finite t. Through the previous inequality, we prove the uniqueness of the solution as follows:
if we assume that in each sub-interval we have at least two solutions   $c_i$ and $\tilde{c}_i$ that satisfy the same initial condition, then for $w_i=c_i-\tilde{c}_i$ we have
\begin{equation}\label{stab5.n}
\xi(t)\leq e^{{\alpha} t}\Big( -2P\int_0^t
 e^{-{\alpha} s}
 \big(w_0(0,s)-w_1(0,s)\big)^2ds\Big).
\end{equation}
This means that $\xi(t) \le 0 $ and consequently $c_i=\tilde{c}_i, i=0,\dots,n.$  The existence of the solution is guaranteed
by showing that an analytic solution can be built as a Fourier series, analogously to
similar diffusion-convective problems in layered systems \cite{pon}.

The upper bound (\ref{stab5})  can be used to describe the behavior of the coupled system for the released mass. Let
\be
   M_i(t)=\int_{x_{i-1}}^{x_i}c_i(s)ds     \qquad M(t)= \sum_{i=0}^n M_i(t)
\ee
be the mass in the layer $i$ and the total mass, respectively. Being $\displaystyle M(t)^2\leq \xi(t) L$ ,  we get, from equation (4.6), the following upper bound for the drug mass in the coupled system
\begin{equation}\label{stab6}
M(t) \leq \sqrt{L}e^{{\alpha \over 2} t}\Big(\xi(0)-2P\int_0^t
 e^{-{\alpha}s}
 \big(c_0(0,s)-c_1(0,s)\big)^2ds\Big)^{1/2}.
\end{equation}

%
%
%The influence of the parameter $P$ in the drug release process can be inferred from the inequality (\ref{stab6}). 
This estimate shows that the larger is $P$, the smaller is the total mass (i.e. the larger is the absorbed mass). The above bound depends on the history of concentration jump at contact interface weighted by controllable quantities such as the applied potential and the media diffusivities.

\subsection{The limit case $P\rightarrow \infty$: perfect contact}\label{subsec::perfect}

We now consider the limit case of $P$, that is when the contact between the vehicle and the target tissue
is perfect, and allows the continuity of the concentration in the interface between
 both media. Differently from the previous subsection, the regularity of the concentration allows the use the
Poincar\'e inequality.  Let $D$, $v$ and $c$ be defined by:
$$D(x)=D_i ,\qquad\quad   v(x)=v_i, \qquad\quad  c(x,t)=c_i(x,t),\qquad \quad   x\in (x_{i-1},x_i),\, i=0,\dots,n.$$
with  (\ref{et1})  replaced by:
\be
J_0 = J_1,\,\quad      c_0 = c_1   \qquad  \mbox{at} \quad x=0,
 \label{et1n}
\ee

Let $(\cdot , \cdot)$ be the usual inner product in $L^2(x_{-1},x_n)$ and let $\| \cdot\|$ the corresponding norm.
We start by observing that we have
\be
\frac{1}{2}\xi'(t)=-(D\nabla c(t),\nabla c(t))-(v c(t),\nabla c(t)) .
 \label{est.1.n}
\ee
Being
$$-(v c(t),\nabla c(t))\leq \max_{i=0,\dots,n}|v_i|\|c(t)\|\|\nabla c(t)\|,$$
and taking into account the following Poincair\'e inequality $\xi(t) \leq \ds\frac{L^2}{2}\|\nabla c(t)\|^2,$  we obtain
$$-(v c(t),\nabla c(t))\leq \max_{i=0,\dots,n}|v_i|\frac{L}{\sqrt{2}} \|\nabla c(t)\|^2.$$
Inserting the last upper bound in (\ref{est.1.n}) we deduce
\be
 \xi'(t) \leq \Big( -2\min_{i=0,\dots,n}D_i +\max_{i=0,\dots,n}|v_i|\sqrt{2} L \Big) \|\nabla c(t)\|^2.
 \label{est.2.n}
\ee
If
\begin{equation}\label{new.cond}
\max_{i=0,\dots,n}\frac{|v_i|}{D_i} L \leq \sqrt{2}.
\end{equation}
by applying the Poincar\'e inequality in (\ref{est.2.n}), we get
\be
 \xi(t) \leq e^{ \beta  t} \|  c(0)\|^2,
 \label{est.4.n}
\ee
where

$$\beta=  \frac{ 1}{L^2}\Big(-2\min_{i=0,\dots,n}D_i +\max_{i=0,\dots,n}|v_i|\sqrt{2} L \Big).$$

Under the condition (\ref{new.cond}) we get the mass  upper bound:
\be
M(t) \leq  \sqrt{L}  e^{\beta   t} \| c(0)\| .
 \label{est.5.n}
\ee

From equation (\ref{est.5.n}), the decreasing of the upper bound for the drug  mass $M(t)$  is observed provided that (\ref{new.cond}) holds. 
%
% Condition (\ref{new.cond}) depends on the relation between the diffusion coeff%icients $D_i$ and the convective velocities $v_i$ that depend on the applied po%tentials $\Psi_0,\Psi_1 $ as well as on the electric conductivity $\sigma_i$.
%The coefficients that arise in the Nernst-Planck relation of the velocity $u_i$  have also role in drug release.

\section{Nondimensional equations}\label{sec::nondimensional}
\setcounter{equation}{0}
 Before solving the differential problem, all the variables, the parameters and the equations are now normalized
to get easily computable nondimensional quantities
as follows:
\begin{eqnarray}
&& \bar x={x \over L}
\qquad\qquad \bar t= { D_{max} \over L^2} t
\qquad\qquad  \bar c_i= {c_i \over C}   \qquad\qquad   \delta={l_0 \over L}
\nonumber \\
&&  \gamma_i={D_i \over  D_{max}}  \qquad\qquad   \bar \phi_i = {F z_i \over RT} \phi_i \qquad \qquad   \Pi={ P L \over D_{max}}   \label{gh2}
\end{eqnarray}
where the subscript $\max$ denotes the maximum value across the $n+1$
layers\footnote{The nondimensional quantity $\ds{F \phi \over RT}= \ds{u \phi \over D}=
\ds{u \nabla \phi L \over D} = {[\mbox{velocity}] \cdot [\mbox{length}] \over [\mbox{diffusivity}]} $
measures the relative strength of electrical driven convective to diffusive forces and corresponds to the P\'eclet number in fluid dynamics problems.  }. By omitting the bar for simplicity,
the 1D  nondimensional Nernst-Planck equations (\ref{er4})--(\ref{es1}) become:
\begin{align}
&{\p c_i \over \p t}=  \gamma_i {\p^2 c_i \over \p x^2} +\gamma_i  \left( {\p c_i \over \p x} { \p \phi_i \over \p x} +c_i { \p^2 \phi_i \over \p x^2} \right) =
\gamma_i {\p^2 c_i \over \p x^2} +\gamma_i a_i  {\p c_i \over \p x}
 \qquad i=0,1,\dots,n\label{eqn1}
\end{align}
(cfr. with (\ref{eyu})).
%The potential is nondimensionalized by scaling  $\phi_i$ (so $a$ and $b$ in
%eqn (3.13)) by $\alpha= \ds{RT \over F}$ ($V$).
The above eqns are supplemented by the following boundary/interface conditions:

\begin{align}
& J_0= -\gamma_0 \left( {\p c_0 \over \p x} + a_0 c_0     \right) =0  \qquad &\mbox{at} \quad x=-\delta \label{ert} \\
& \nonumber \\
& J_0 =- \gamma_0 \left(  {\p c_0 \over \p x} + a_0 c_0   \right)=
 -\gamma_1 \left(  {\p c_1 \over \p x} + a c_1  \right) =J_1
  \nonumber  \\
& J_0= -\gamma_0 \left(  {\p c_0 \over \p x} + a_0 c_0   \right)=\Pi (c_0- c_1)
\qquad &\mbox{at} \quad x=0 \label{bc0} \\
&\nonumber \\
& J_i =- \gamma_i \left(  {\p c_i \over \p x} + a c_i  \right)=
 -\gamma_{i+1} \left(  {\p c_{i+1} \over \p x} + a c_{i+1}   \right) =J_{i+1}  \nonumber \\
& c_i =c_{i+1} \qquad i=1,\dots,n-1 \qquad &\mbox{at} \quad x=x_i \label{bc3} \\
&\nonumber \\
& c_n=  0 \qquad &\mbox{at} \quad x=1  \label{bc1}
\end{align}
%Note that from equation (13) (flux being sum of 2 positive quantities) it
%follows that $c_0=  \ds{\p c_0 \over \p x}=0 $.
and initial conditions:

\be
c_0(x,0)=1  \qquad\qquad\qquad c_i(x,0)=0  \quad i=1,2,\dots,n
\ee

\section{Numerical solution}\label{sec::FD_scheme}
\setcounter{equation}{0}
Although a semi-analytic treatment is possible in multi-layered diffusion problems \cite{pon2}, we proceed to solve the nondimensional system of equations (\ref{eqn1})--(\ref{bc1}) numerically.
Let us subdivide the interval $(-\delta,0)$ into  $N_0+1$ equispaced
grid nodes $x_0^j=(j-N_0) \, h_0 ,\quad j=0,1,\dots,N_0$,  and the $i-$th interval $[x_i,  x_{i+1}]$ with $N_i+1$ equispaced  points $x_i^j=j \, h_i ,\quad j=0,1,\dots,N_i$. Here, $h_0,h_1, h_2\dots,h_n$ represent the spacing in the vehicle (layer 0) and skin (layers $i=1,2\dots,n$), respectively\footnote{In the following, the subscript $i$ refers to  the layer, the superscript $j$ denotes the approximated value of the concentrations  at $x_i^j$.}.
In each layer, we approximate the diffusive terms by considering a standard finite-difference of the second derivative and centered first derivative at internal nodes $x_j$:
\begin{align}
& \left. { \p c_i \over \p x} \right|_{x_j} \simeq  {c_i^{j+1} -  c_i^{j-1}  \over 2  h_i}    \nonumber  \\
&  \label{ev1}    \\
& \left. { \p^2 c_i \over \p x^2} \right|_{x_j} \simeq  { c_i^{j-1} - 2 c_i^{j} +c_i^{j+1} \over h_i^2}   \qquad\qquad  j=1,\dots,N_i-1 \qquad  i=0,1,\dots,n   \nonumber  
\end{align}

\bigskip

At  the boundary points $x=-\delta$ and  $x=1$  and at interfaces $x_i$, 
the equations
 (\ref{eqn1}) hold,
but the approximations (\ref{ev1}) include the boundary conditions (\ref{ert})--(\ref{bc1}) and the interface conditions (\ref{bc3}).

\bigskip

\noindent\underline{Treatment of the interface $x=0$} \\
At the interface  $x=0$, we potentially have a discontinuity in concentration and two possibly different values, say  $\tilde c_0^N$ and $\tilde c_1^0$ (the tilde accent indicates these special points) , each for each interface side, need  to be determined (fig. 4).

 No derivative can be computed across the interface $x=0$,
due to a possible
discontinuity and approximations (\ref{ev1})  no longer apply. An alternative procedure is needed in eqn (\ref{ev1}) to get $\tilde c_0^N$  for $j=N_0-1$ and $\tilde c_1^0$  for $j=1$.  Their values are related through the
 interface conditions (\ref{bc0}):
\begin{align}
  & - \gamma_0 \left( {\p  \tilde c_0^N \over \p x} + a_0 \tilde c_0^N \right) = - \gamma_1 \left( {\p  \tilde c_1^0 \over \p x} + a \tilde c_1^0 \right)  \nonumber \\
& - \gamma_0 \left( {\p \tilde c_0^N \over \p x} +a_0 \tilde c_0^N \right)= \Pi (\tilde c_0^N -  \tilde c_1^0) \label{ap2}
\end{align}
Following the approach described by Hickson et al. \cite{hick} , we take a Taylor series expansion for  $c_0^{N-2} , c_0^{N-1},  c_1^{1} , c_1^{2}$, and arrive at:
\begin{align}
&c_0^{N-1} \approx \tilde c_0^N - h_0 {\p \tilde c_0^N \over \p x} + {h_0^2 \over 2} {\p^2 \tilde c_0^N \over \p x^2} \nonumber \\
&c_0^{N-2} \approx \tilde c_0^N - 2 h_0 {\p \tilde c_0^N \over \p x} + 2 {h_0^2 } {\p^2 \tilde c_0^N \over \p x^2}  \nonumber \\
&c_1^{1} \approx \tilde c_1^0 + h_1 {\p \tilde c_1^0 \over \p x} + {h_1^2 \over 2} {\p^2 \tilde c_1^0 \over \p x^2} \nonumber \\
&c_1^2 \approx \tilde c_1^0 + 2 h_1 {\p \tilde c_1^0 \over \p x} + 2 {h_1^2} {\p^2 \tilde c_1^0 \over \p x^2}  \label{ap1}
\end{align}
The two equations (\ref{ap2}) and the four equations (\ref{ap1}) form an algebraic system of six equations that allow to express $\tilde c_0^{N}, \tilde c_1^0 $ and their first and second derivatives as a linear
combination of the neighboring values. Using symbolic calculus  we obtain:

\begin{align}
\tilde c_0^N= {[\gamma_0 \gamma_1 (3-2 a h_1) + 2 \Pi \gamma_0 h_1 ] ( c_0^{N-2} - 4 c_0^{N-1} )  + 2 \gamma_1 h_0 \Pi (c_1^{2} - 4 c_1^{1}) \over Q}  \label{tr4} \\
\tilde c_1^0= { [\gamma_0 \gamma_1 (3+2 a_0 h_0) + 2 \Pi \gamma_1 h_0 ] ( c_1^{2} -  4 c_1^{1} )  + 2 \gamma_0 h_1 \Pi (c_0^{N-2} - 4 c_0^{N-1})
 \over Q}  \label{tr5}
\end{align}
where $Q= 2 \gamma_1 h_0 \Pi (2 a h_1 -3) +\gamma_0 (2 a_0 h_0 +3)(2 a \gamma_1  h_1- 3 \gamma_1 -2 h_1 \Pi )$.
\\

\noindent After spatial discretization, the system of PDEs reduces to a system of nonlinear ordinary differential equations (ODEs) of the form:
\be
{dY \over dt} = A ( Y)  \label{ew1}
\ee
where $Y=(c_0^0 ,\dots.   ,c_0^{N-1} , c_1^1 , \dots   ,c_1^{N_1} ,\dots., c_n^0, \dots.   c_n^{N_n} )^T$
and $A(Y)$ contains the coupled $N_0+N_1+\dots+N_n$ discretized equations  (\ref{eqn1}).
The system (\ref{ew1}) is solved by the routine ode15s of {\tt Matlab} based on a Runge-Kutta type method with backward differentiation formulas, and an adaptive time step.
The interface drug concentrations $\tilde c_0^{N}, \tilde c_1^0 $  are computed
a posteriori through equations (\ref{tr4})--(\ref{tr5}).

\section{Results and discussion}\label{sec::numerical_results}
\setcounter{equation}{0}
A common difficulty in simulating physiological 
processes, in particular TDD,
is the identification of reliable estimates of the model parameters.
Experiments of TDD are impossible or
prohibitively expensive in vivo and the only available source
are lacking and incomplete data from literature.
The drug delivery problem depends on a large number of constants,
each of them varies in a finite range, with  a variety of combinations and limiting cases.
Furthermore, these parameters can be influenced
by body delivering site,  patient age and individual variability.
They cannot be chosen independently from each other and there is a compatibility condition among them.
%In this paper the following constants are used s in TDD %\cite{pra}
%%\cite{ana,pon2,kub1,sim1}:
%\begin{align}
% & P= 10^{-5} cm /s   \qquad   \Delta \Psi= \Psi_0 %-\Psi_1= 1 \div 10  V  \nonumber \\%%
%&   F=96485.33 \, C mol^{-1}  \qquad   R=  8.31 J K^{-1} mol^{-1}  \qquad  T=300 K   \label{eqr}
%\end{align}
 
Here, the skin is assumed to be composed of three main layers, say the
stratum corneum, the viable  epidermis, and the dermis with respective model parameters given in table 1. In the absence of direct measurements, indirect data are inferred from previous studies in literature  \cite{mill,pra,becker}.
Diffusivities critically depend on the kind and size of the transported molecules and are affected of a high degree of uncertainty. The vehicle-skin permeability parameter is estimated as $P= 10^{-5} cm /s $.

\begin{table}[ht]
\caption{The parameters used in the simulations for the vehicle and the three skin layers.}
\centering
\begin{tabular}{|| c | c | c | c | c ||}
\hline
 \textemdash  & vehicle (0) & stratum corneum (1) & viable epidermis (2) & dermis (3)   \\ \hline
$l_i=x_i-x_{i-1}(cm)$ &  $ 0.01$  & $1.75 \cdot 10^{-3}$ &  $3.5 \cdot 10^{-3}$   & $0.11$  \\ \hline
$D_i (cm^2/s)$ &   $10^{-4}$ & $10^{-10}$ &  $ 10^{-7}$ &  $10^{-7}$  \\ \hline
$\sigma_i (S/cm)$ &   $1.5 \cdot 10^{-2}$  & $10^{-7}$ & $10^{-4}$   & $10^{-4}$  \\ \hline
\end{tabular}
\end{table}

The coefficients of potential $\phi_i$ in eqn (\ref{eyu})  have the following expressions:

\begin{align}
& a_0= -{\Delta \Psi \sigma_1 \sigma_2 \sigma_3  \over G } \nonumber \\
& b_0= {l_0 \Psi_1 \sigma_1 \sigma_2 \sigma_3 +l_1 \Psi_0 \sigma_0 \sigma_3 (\sigma_2 - \sigma_1)+
l_2 \Psi_0 \sigma_0 \sigma_1 ( \sigma_3 -  \sigma_2) + l_3 \Psi_0 \sigma_0 \sigma_1 \sigma_2 \over G}  \nonumber \\
&a_1=  -{\Delta \Psi \sigma_0 \sigma_2 \sigma_3  \over G} \nonumber \\
& b_1={l_0 \Psi_1 \sigma_1 \sigma_2 \sigma_3 +l_1 \Psi_0 \sigma_0 \sigma_3 (\sigma_2 - \sigma_1)+
l_2 \Psi_0 \sigma_0 \sigma_1 ( \sigma_3 -  \sigma_2) + l_3 \Psi_0 \sigma_0 \sigma_1 \sigma_2 \over G} =b_0  \nonumber \\
&a_2=  -{\Delta \Psi \sigma_0 \sigma_1 \sigma_3  \over G } \nonumber \\
&b_2= {l_0 \Psi_1 \sigma_1 \sigma_2 \sigma_3 +l_1 \Psi_1 \sigma_0 \sigma_3 (\sigma_2 - \sigma_1)+
l_2 \Psi_0 \sigma_0 \sigma_1 ( \sigma_3 -  \sigma_2) + l_3 \Psi_0 \sigma_0 \sigma_1 \sigma_2 \over G}    \nonumber \\
&a_3=  -{\Delta \Psi \sigma_0 \sigma_1 \sigma_2  \over G } \nonumber \\
&b_3=  {l_0 \Psi_1 \sigma_1 \sigma_2 \sigma_3 +l_1 \Psi_1 \sigma_0 \sigma_3 (\sigma_2 - \sigma_1)+
l_2 \Psi_1 \sigma_0 \sigma_1 ( \sigma_3 -  \sigma_2) + l_3 \Psi_0 \sigma_0 \sigma_1 \sigma_2 \over G}
\end{align}
where $G=l_0 \sigma_1 \sigma_2 \sigma_3 + l_1 \sigma_0\sigma_3 (\sigma_2- \sigma_1)
+ l_2 \sigma_0\sigma_1 (\sigma_3- \sigma_2) +l_3 \sigma_0 \sigma_1 \sigma_2$ and
$\Delta \Psi= \Psi_0 - \Psi_1$.
Note that only the electric potential gradients, $a_i$, appear in the Nernst-Planck equation and the ratio of the slopes satisfies: $\ds{a_i \over a_{i+1}}=\ds{\sigma_{i+1} \over \sigma_i}$.
%and $b_i$ are never used in the model. \par

Instead than reproducing the clinical protocols where iontophoresis consists 
in repeated sessions of 10-20 mins, the numerical simulations aim at bringing to extreme values  both 
 the potential (up to $\Delta \Psi= 10V$) 
and the duration of application (30 mins): a current is activated during this period of time and 
then switched off.
In fig. 5 the concentration profiles are shown before (continuous lines) and after (dashed lines) the current application. The iontophoretic effect is  to enhance the stratum corneum permeation but, only for higher $\Delta \Psi$, appears to be significant in  deeper layers.  No relevant effect is present after suspension.  It is desirable  the drug level is  maintained over a certain amount to exerts its therapeutic effect without exceeding a given threshold to not be toxic. \par

In fig. 6 the different distribution of drug mass in all layers is depicted:
mass is decreasing in the vehicle, whereas in the other layers is  first increasing, reaches a peak, and asymptotically decreasing, as in other similar drug delivery systems  \cite{pon, pon2}. Due to the sink condition at the right end, part of mass is lost via the systemic circulation. By considering this effect, a  drug mass 
conservation holds  and the  progressive emptying of the vehicle 
corresponds to the
drug replenishment of the other layers -- in a cascade sequence  -- at a rate depending of
the electric field (fig. 6).
Again, an augmented transdermal permeation is reported with higher values of the potential: this is more effective during current administration, but prolongs at later times ($12h-72h$).  The transport of a species across skin will be
determined by the strength and the duration of the electric field, the concentration and the mobility of all ions in the skin. 
The desired delivery rate is obtained with a proper choice of the
physico-chemical-electrical parameters.  \par
These outcomes provide valuable indications to assess whether
drug reaches a deeper layer, and to optimize the dose capacity in
the vehicle.  Therefore, it is possible to identify the conditions 
that guarantee
 a more prolonged and uniform release or a localized peaked
distribution.

\section{Conclusions}\label{sec::conclusions}

Nowadays  iontophoretic systems are commonly used
 in transdermal applications.
These systems use  an electric field to enhance the release from a
drug reservoir and to direct 
the therapeutic action at the target tissue 
with a given rate and at desired level.  Iontophoresis provides 
a mechanism to control the transport of
hydrophilic and charged agents across the skin, especially 
for high molecular weight substances 
such as peptides or proteins which are
usually ionized and hardly penetrate the stratum corneum  by
conventional passive diffusion. Notwithstanding, 
the effective utilization of electric field-assisted 
transport for drug delivery across biological
membranes  requires a deeper understanding of the mechanisms and
theories behind the process.  \par
In this paper a   multi-layer model is
developed to clarify the role of the applied potential, 
the conductivity of the skin, the drug diffusion
and the systemic absorption. 
The stability of the mathematical  problem is discussed within two different
scenarios: imperfect and perfect contact between the reservoir and the
target tissue. To illustrate the drug dynamics in the composite medium - vehicle and skin layers coupled - during  and after the electric 
administration,
an accurate finite-difference method is proposed.  
  The  modelling approach allows the simulation in several experimental setting, including extreme conditions which are not possible in
clinical environment. Numerical experiments show to what extent
the applied current, along the duration of application,
accelerates the  depletion of the reservoir and increases  the drug absorption in the deep skin.  The present TDD model constitutes a simple and useful tool 
in exploring new delivering strategies that guarantee the
optimal and localized release for an extended period of time.

\section*{Acknowledgments}
 This work was partially supported 
by the Centre for Mathematics of the
University of Coimbra -- UID/MAT/00324/2013, funded by the Portuguese
Government through FCT/MEC and co-funded by the European Regional Development Fund through the Partnership Agreement PT2020. The support of the bilateral project FCT-CNR 2015-2016 is greatly acknowledged.  \par
We are grateful to E. Di Costanzo for many valuable
discussions and helpful comments.

\newpage

\begin{figure}[hb] 
\centering
\scalebox{0.15}{\includegraphics{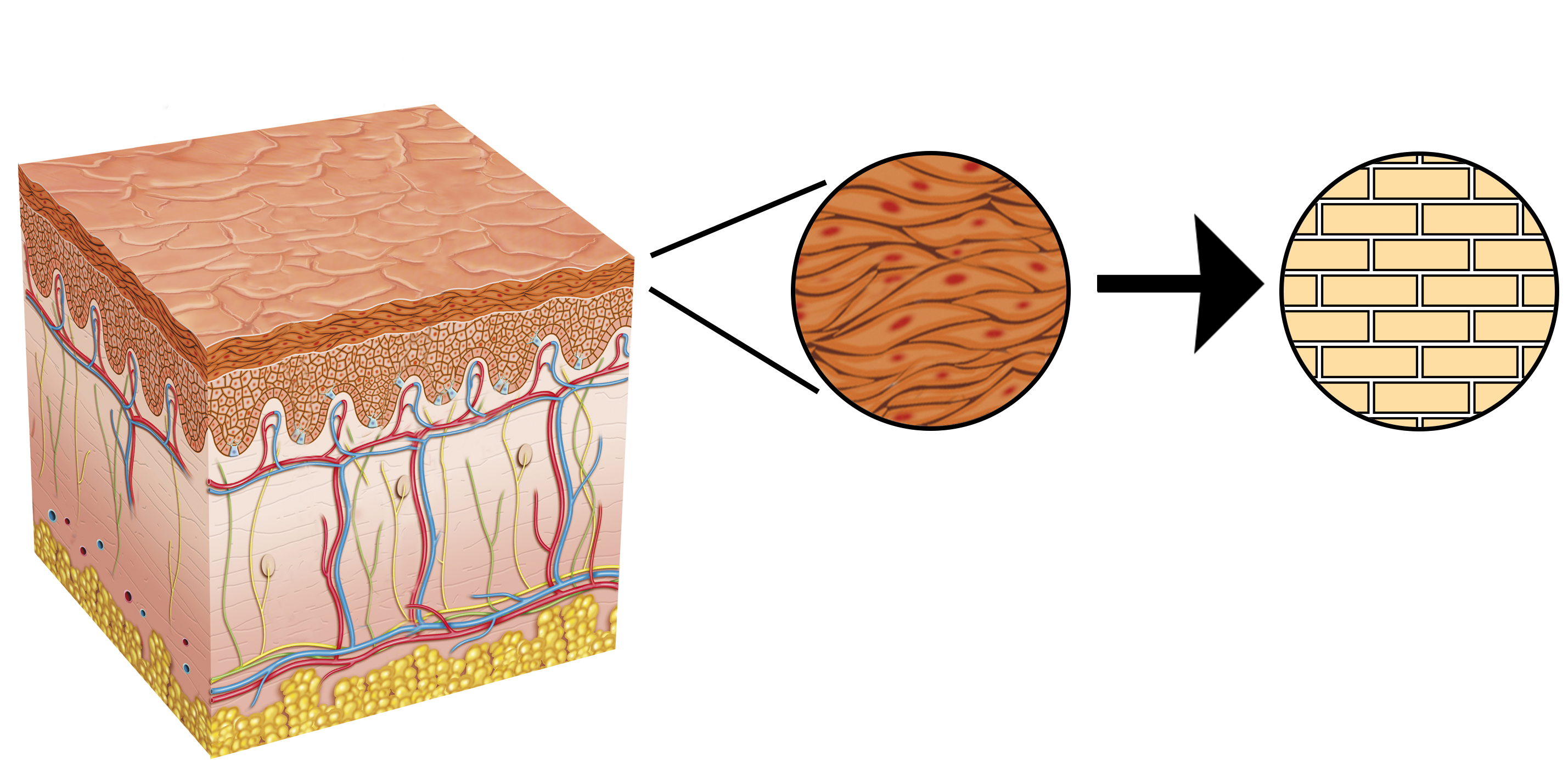}} 
\caption{An anatomic representation of the skin,
 composed by three main layers: epidermis (approximately $100  \mu m$ thick), dermis (between $1$ and $3mm$ thick, highly vascularized) and a subcutaneous tissue. The epidermis is divided into sub-layers where the  stratum corneum (approximately $15 \mu m$ thick) is the outermost layer and is the major   barrier to the drug migration, being composed of densely packed cells, with a typical {\em brick and  mortar} structure. Each skin layer, due to its histological composition, has a different influence in the drug transport mechanism.  }  
\end{figure}
%\end{psfrags}

\begin{figure}[hb] 
\centering
\scalebox{0.8}{\includegraphics{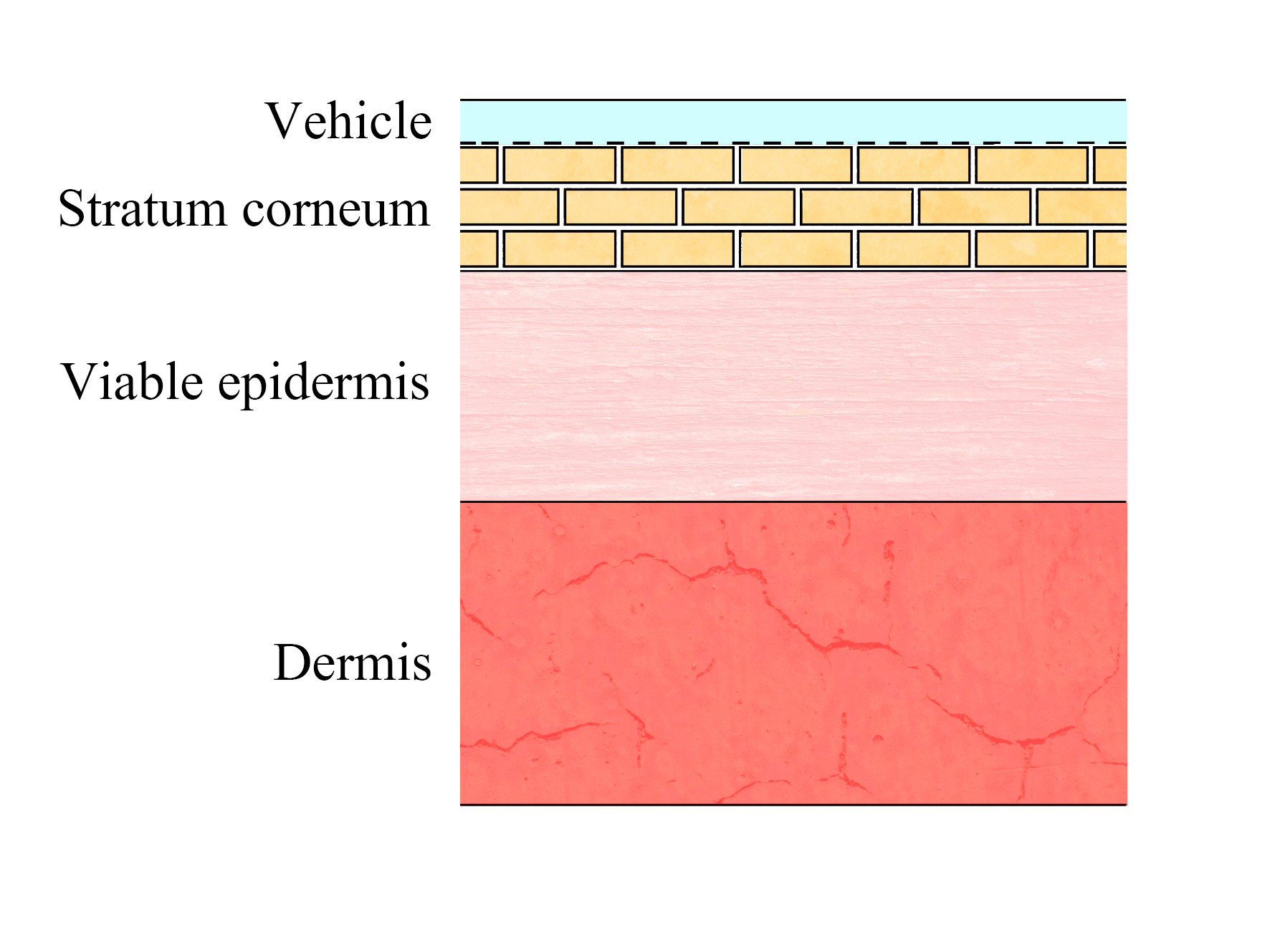}} 
\caption{A schematic section
representing  the present  multi-layered model. A vehicle is applied over the skin surface with an imperfect contact  at the vehicle-skin
interface (figure not to scale). }  
\end{figure}
%\end{psfrags}

\begin{figure}[hb] 
\centering\scalebox{0.8}{\includegraphics{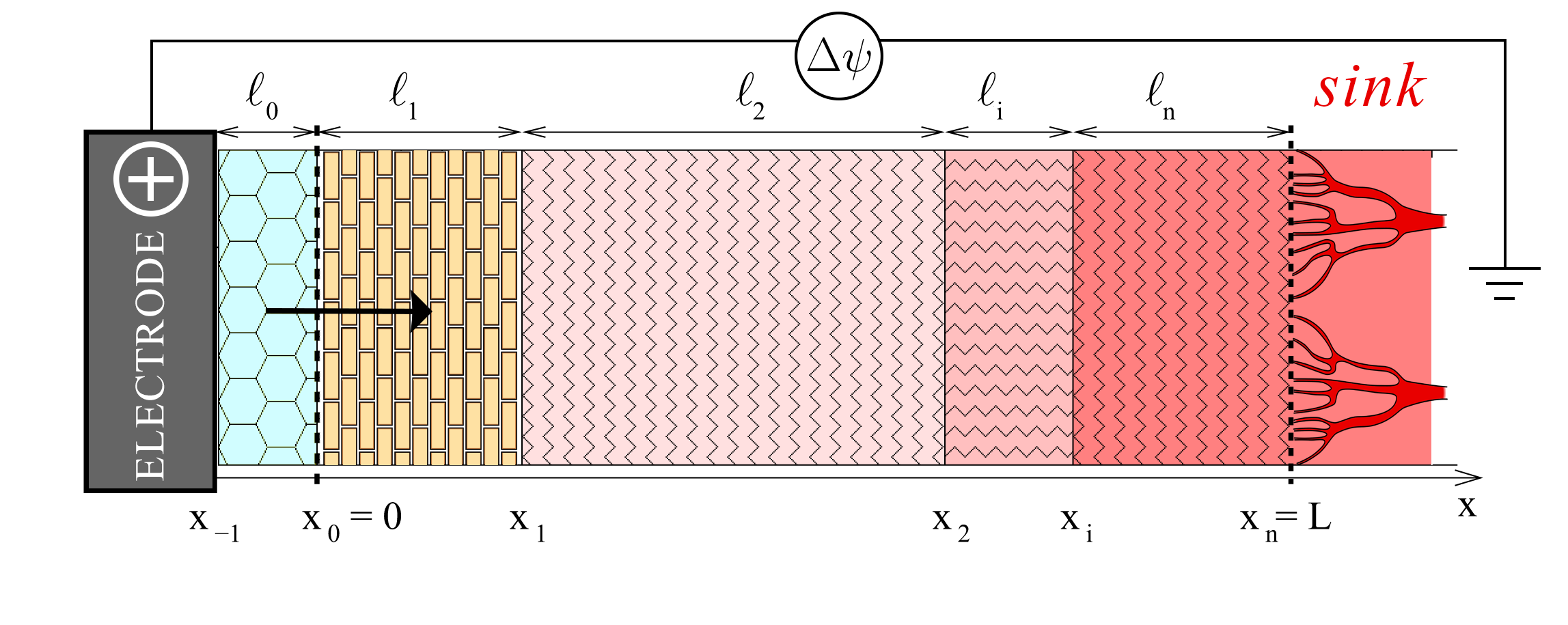}} 
\caption{A diagram sketching a $n$-layered tissue ($\ell_1, \ell_2,...,\ell_n$) 
faced with the vehicle $\ell_0$. 
The 1D model is defined along the line normal to the skin surface and extends
with a sequence of $n$ contiguous layers  from the vehicle interface
$x_0=0$ up to the skin bound $x_n=L$, where capillaries sweep the drug away to the systemic circulation (sink). In iontophoresis, a difference of potential is applied to facilitate  drug penetration 
from the vehicle across the tissue's layers
(figure not to scale).}  
\end{figure}
%\end{psfrags}

\begin{figure}[ht!]
\centering
\scalebox{0.8}{\includegraphics{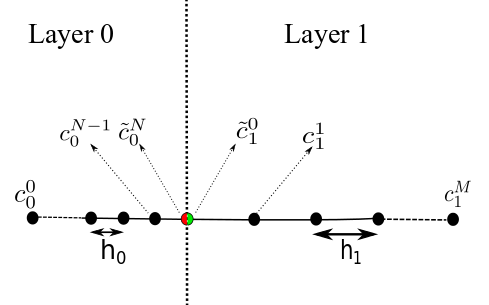}}
\caption{Schematic illustration of grid nodes in the first
 two layers and the interface
points (in red and green).
They are computed a posteriori as a linear combination of the
neighbouring grid points (eqns. (\ref{tr4})--(\ref{tr5})). At the other
interfaces a single concentration values occur.  }
\end{figure}
\begin{landscape}
\begin{figure}
%\begin{turn}{-90}
%\begin{centering}
\begin{tabular}[t]{ccccc}
 & $c_0$ & $c_1$ & $c_2$ &  $c_3$ \tabularnewline
\parbox[t]{1.86cm}{\vspace{-2.2 cm}$\Delta\Psi=0V$}
&\includegraphics[scale=0.25]{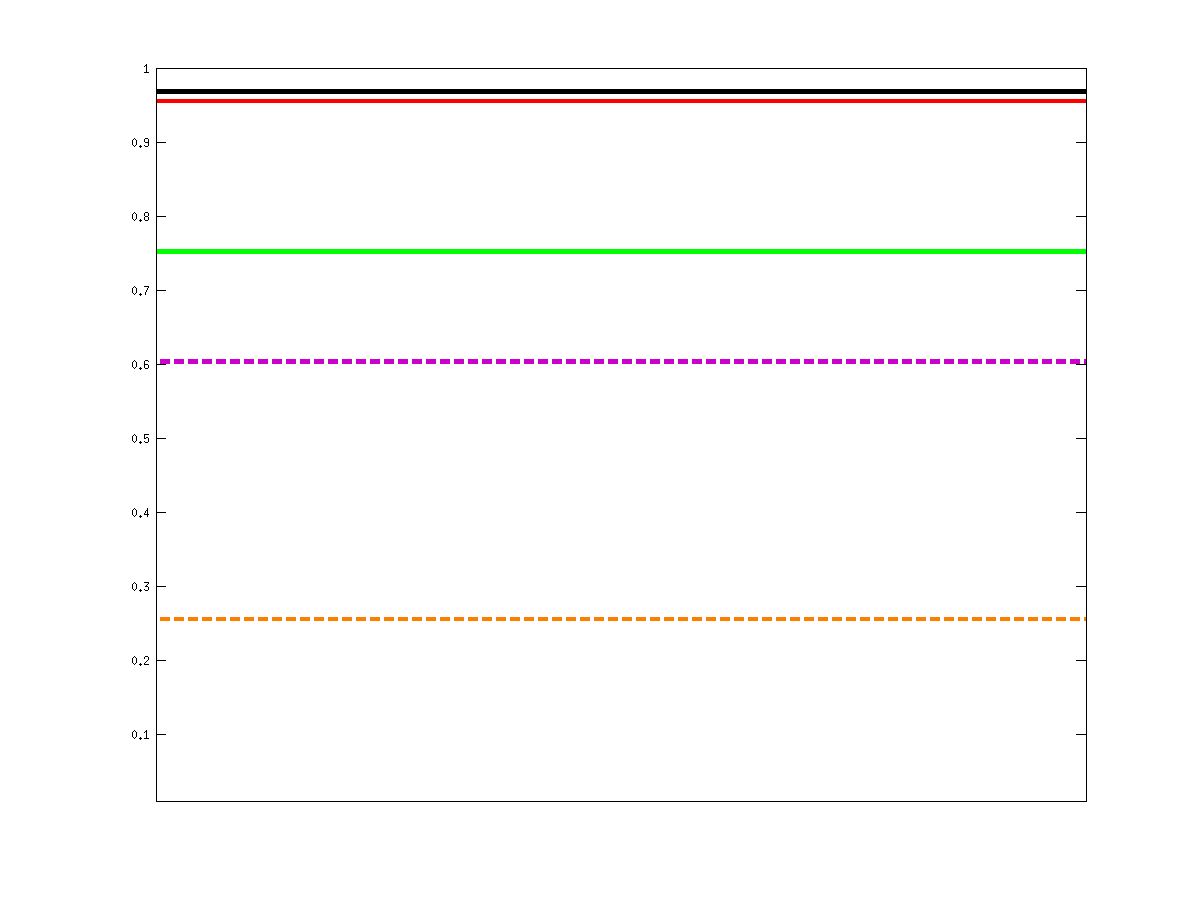}
&\includegraphics[scale=0.25]{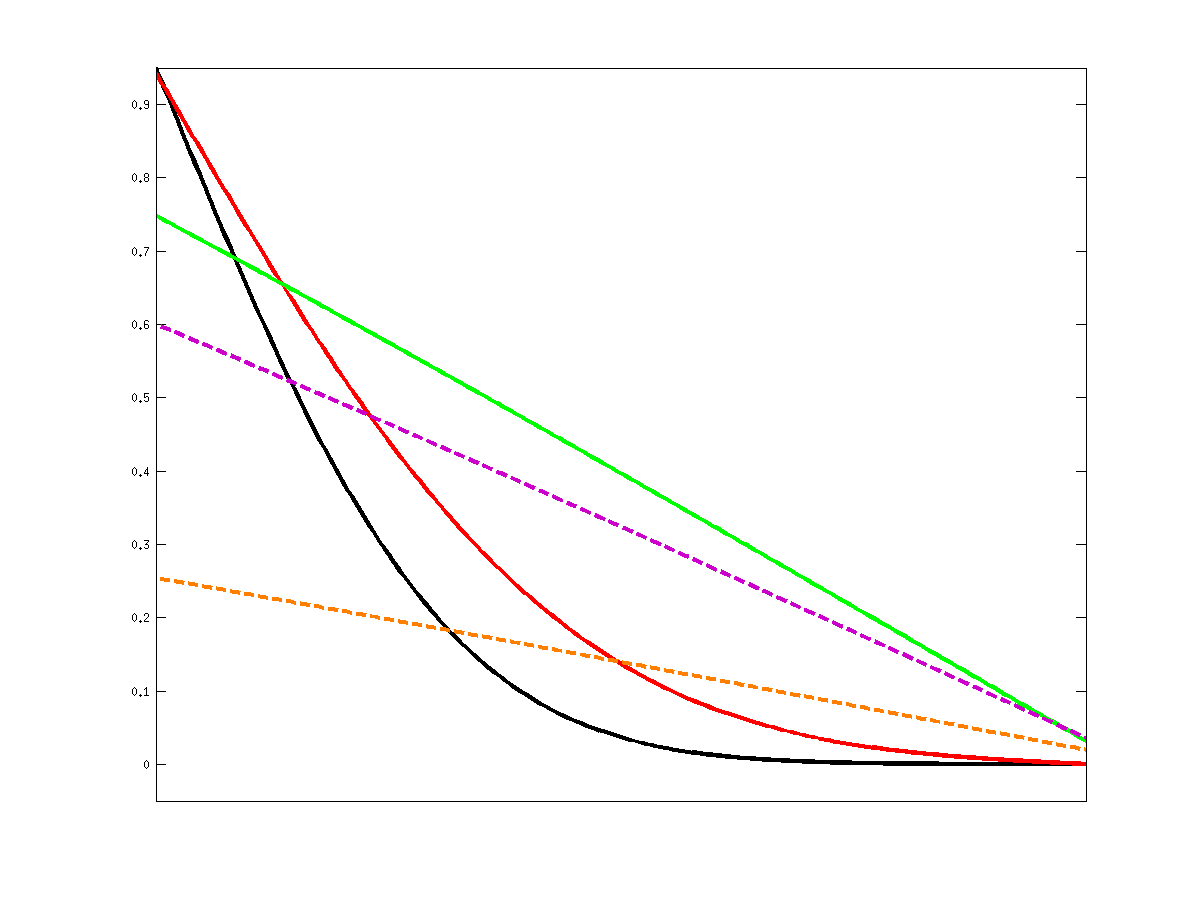}
&\includegraphics[scale=0.25]{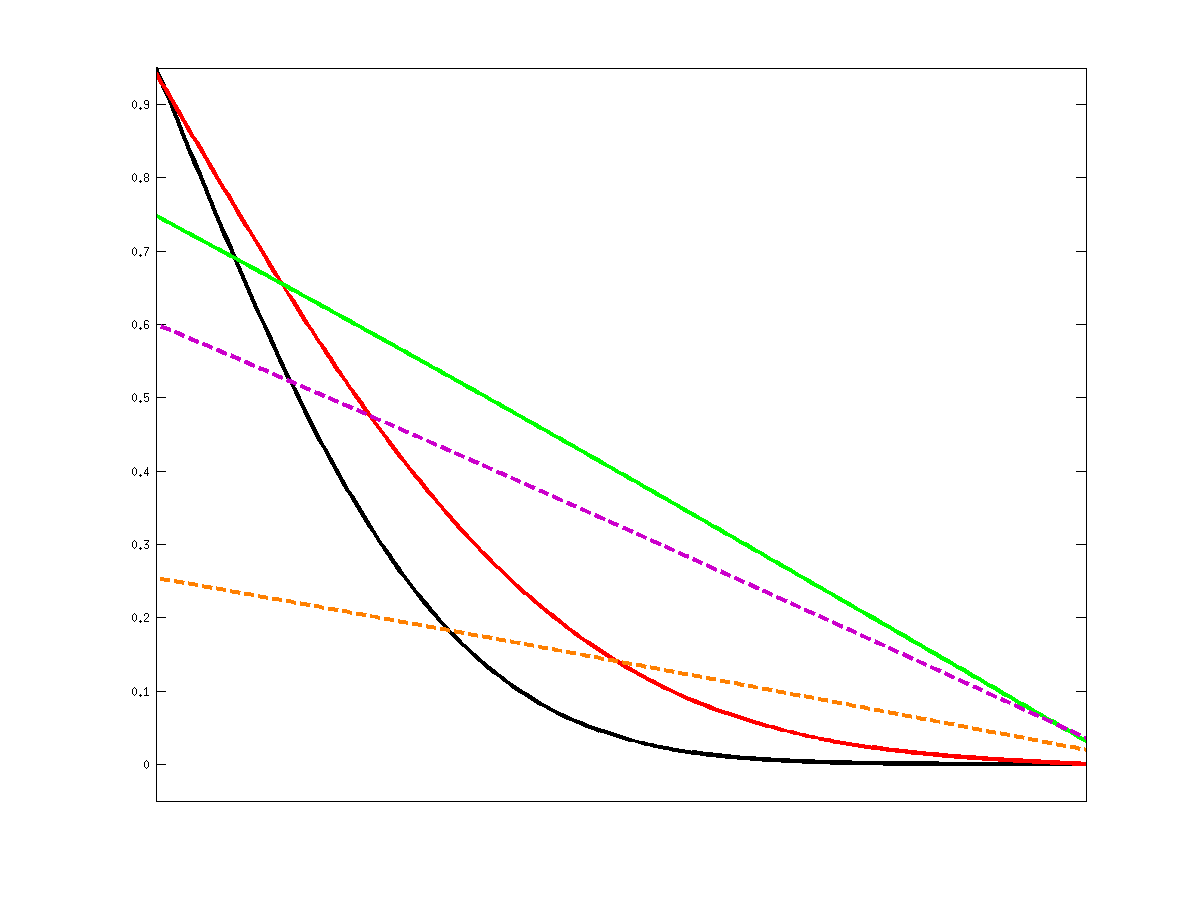}
&\includegraphics[scale=0.25]{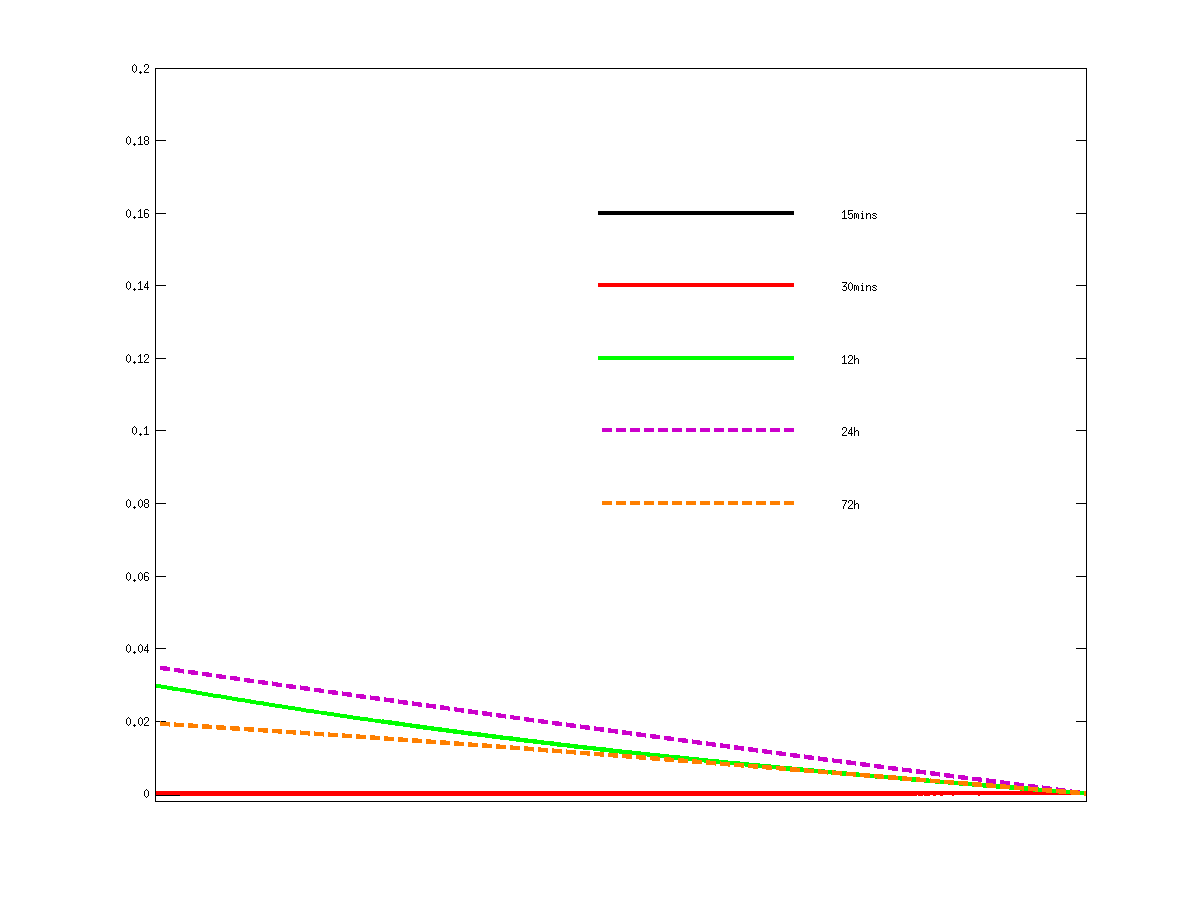}
\tabularnewline
\parbox[t]{1.8cm}{\vspace{-2.2 cm}$\Delta\Psi=1V$}
&\includegraphics[scale=0.25]{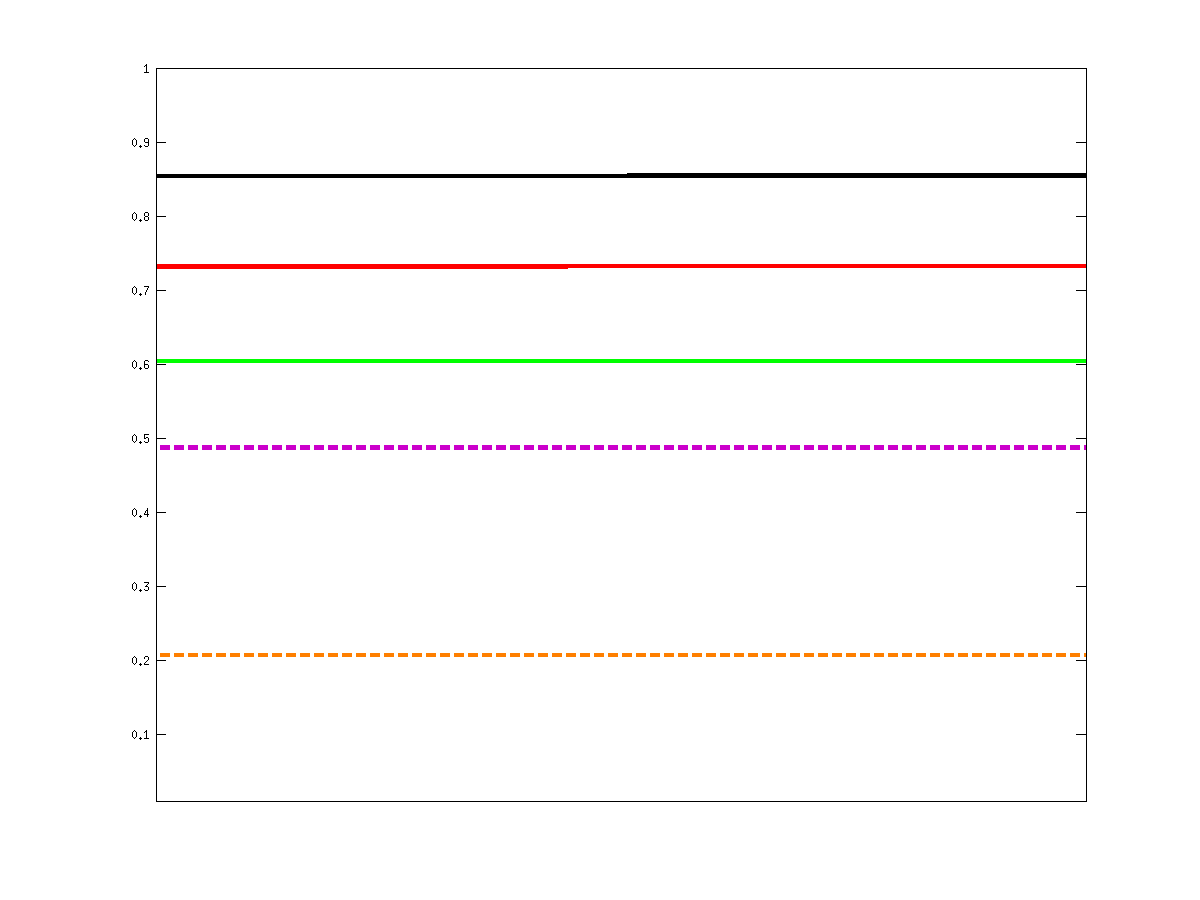}
&\includegraphics[scale=0.25]{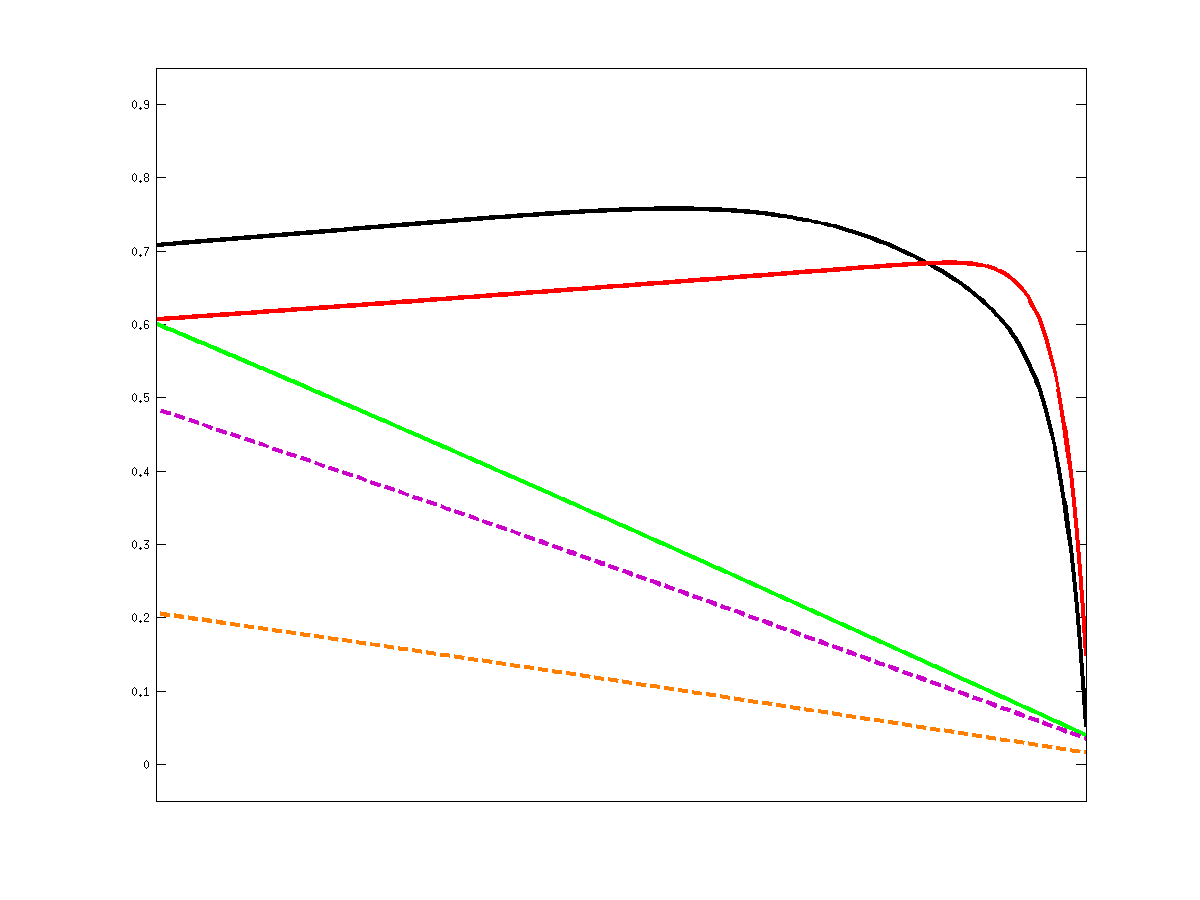}
&\includegraphics[scale=0.25]{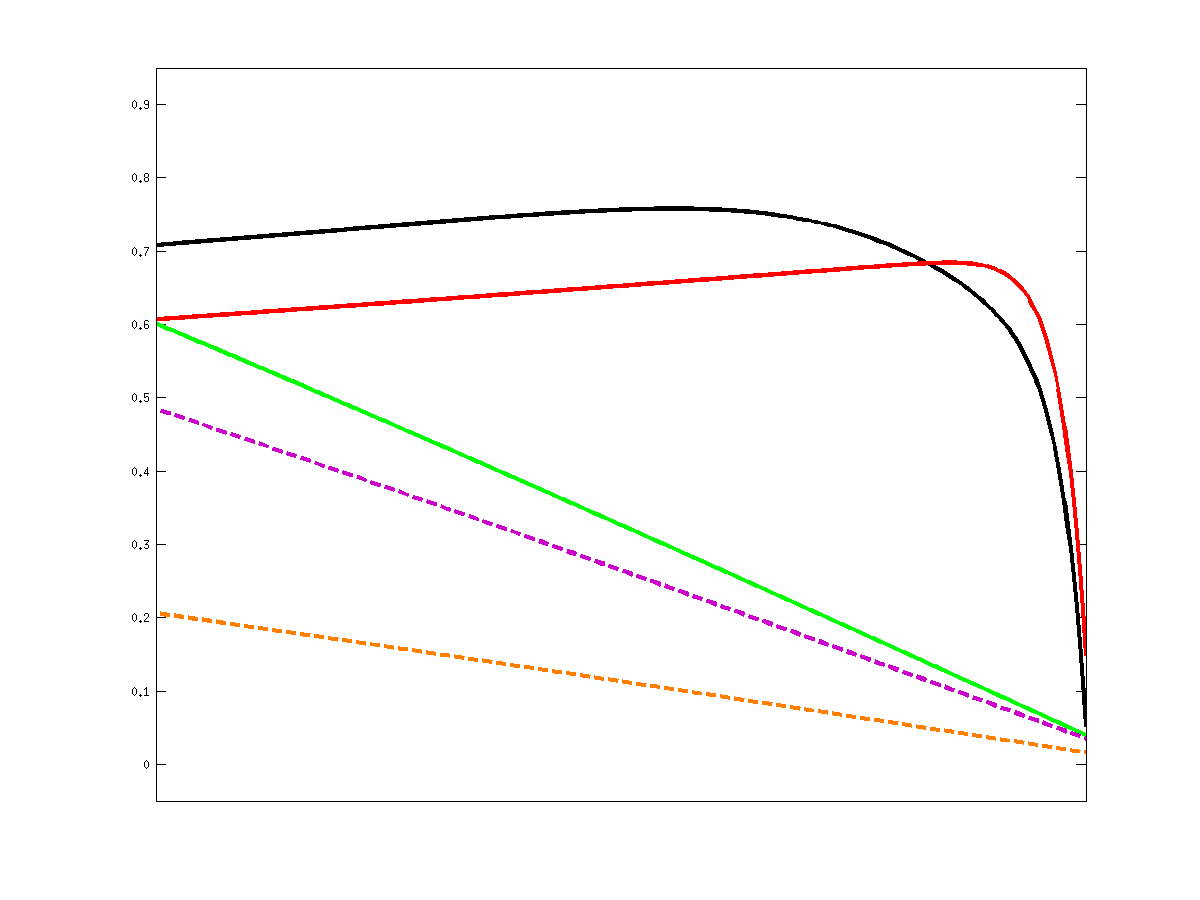}
&\includegraphics[scale=0.25]{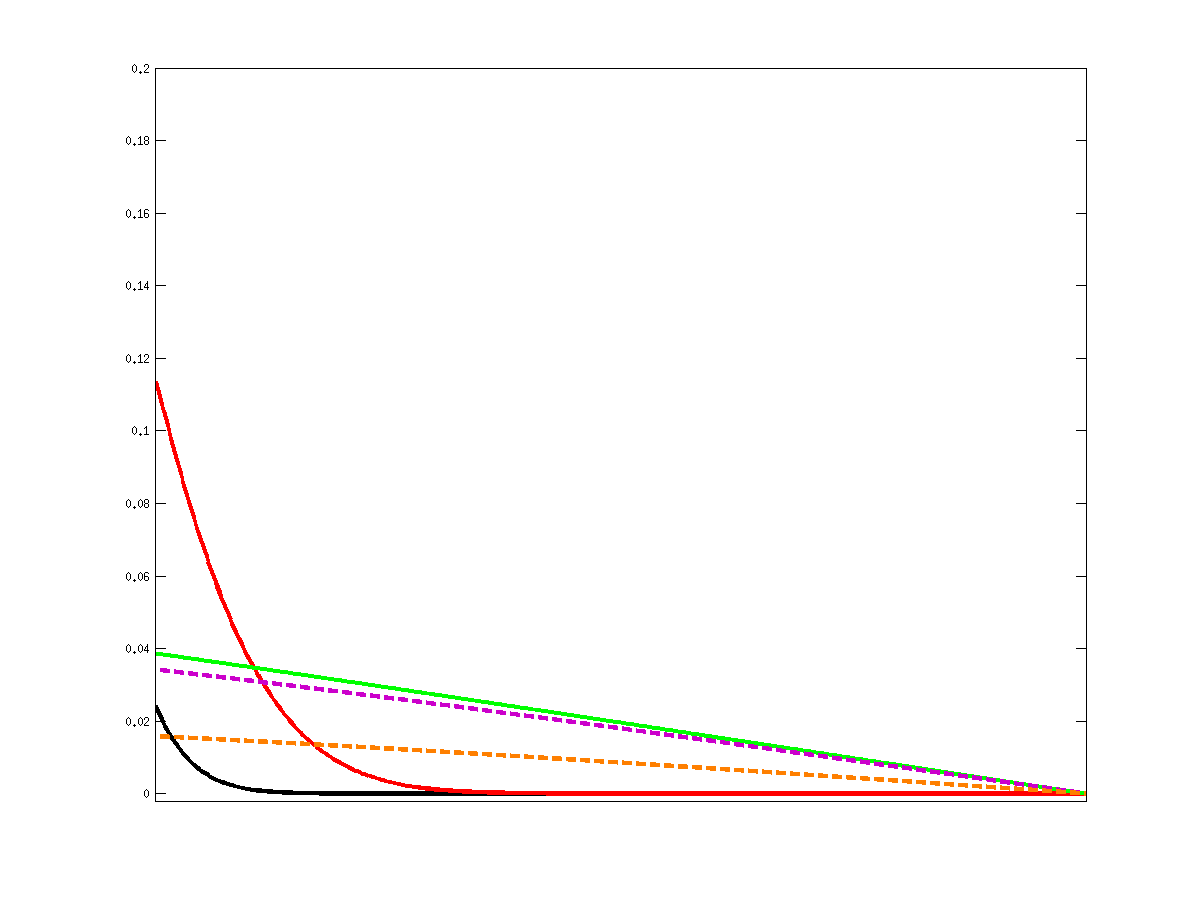}
\tabularnewline
\parbox[t]{2.2cm}{\vspace{-2.2 cm}$\Delta\Psi=10V$}
&\includegraphics[scale=0.25]{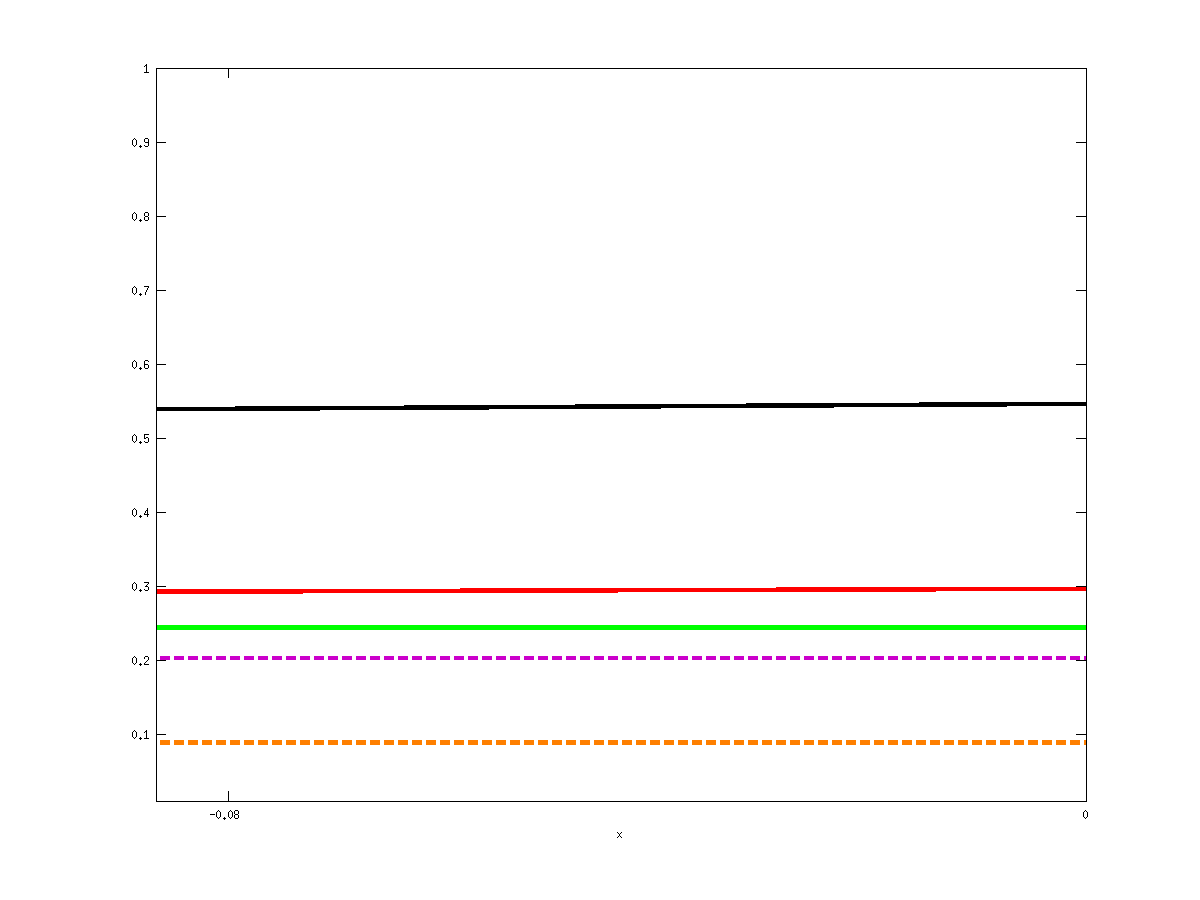}
&\includegraphics[scale=0.25]{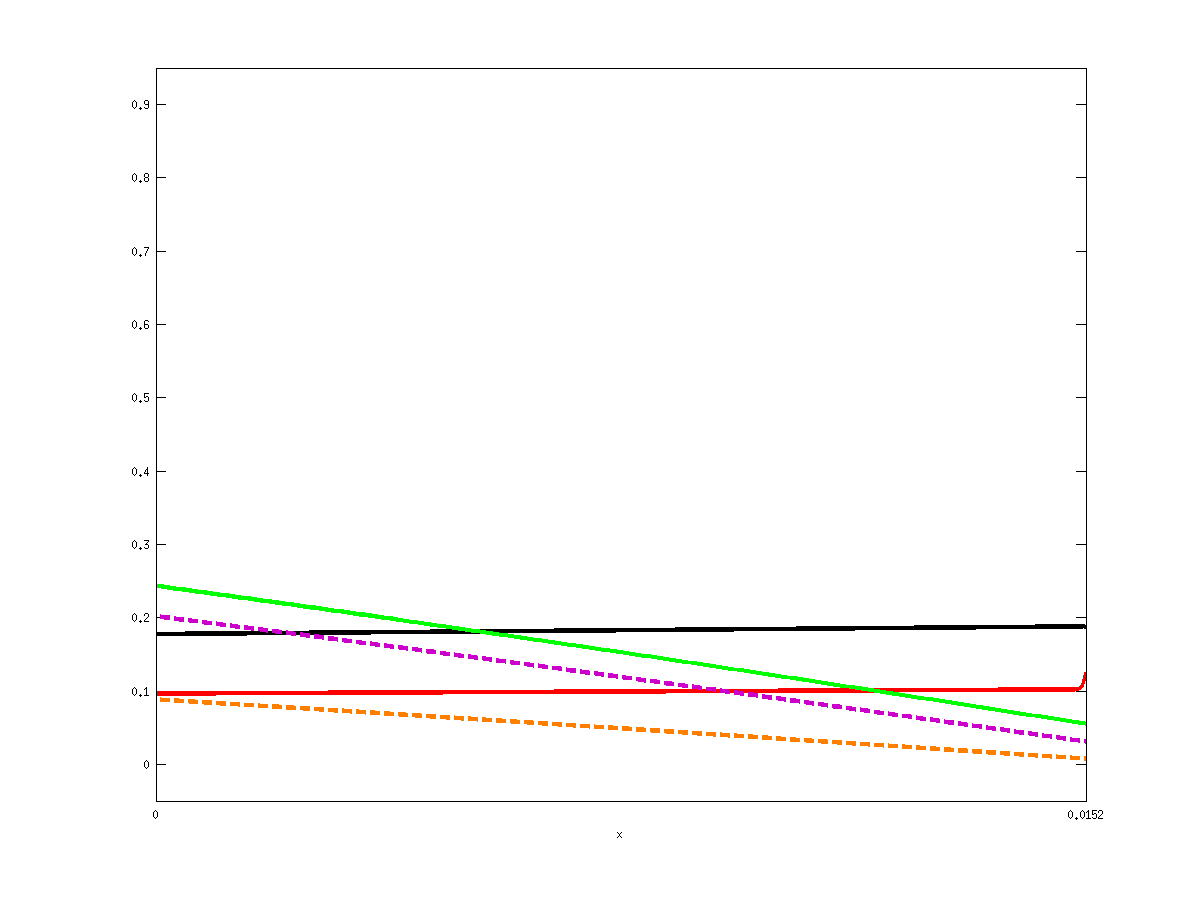}
&\includegraphics[scale=0.25]{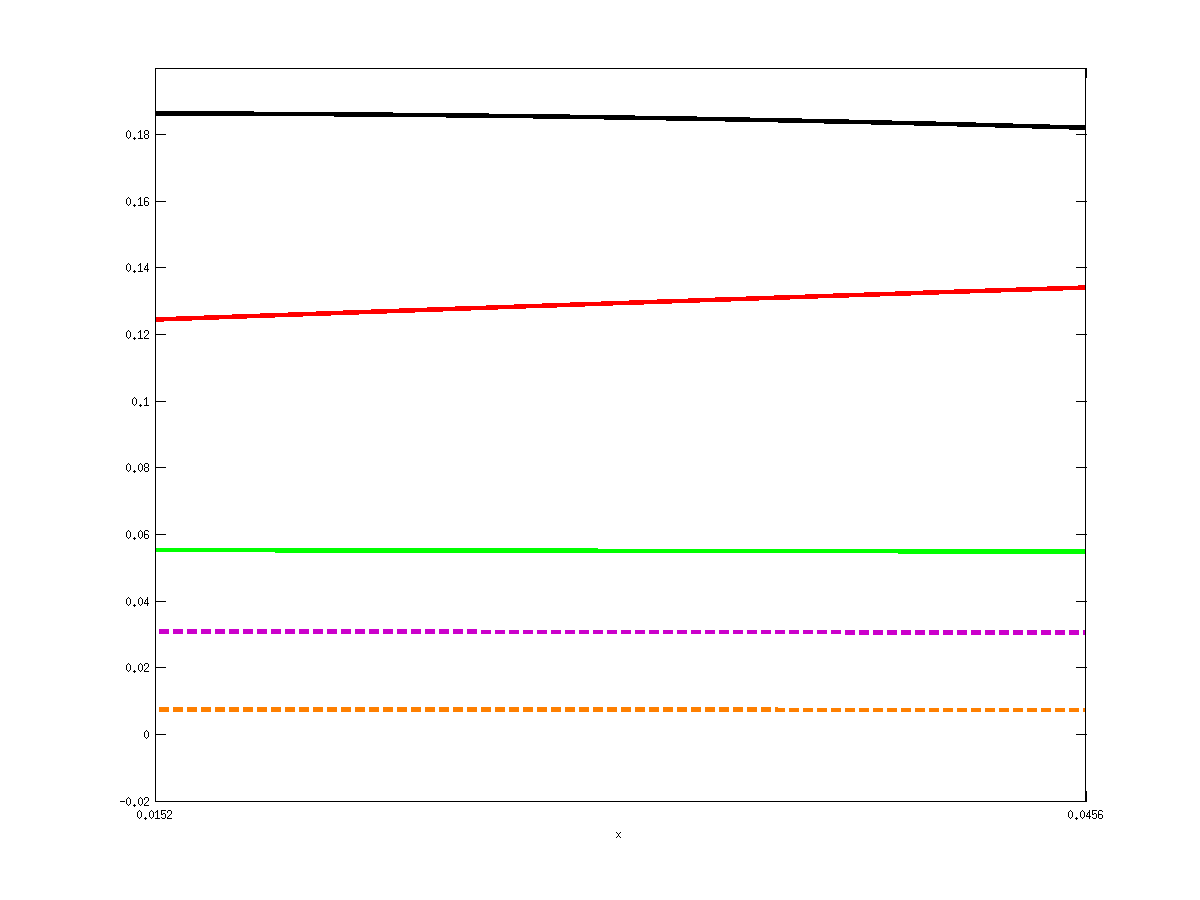}
&\includegraphics[scale=0.25]{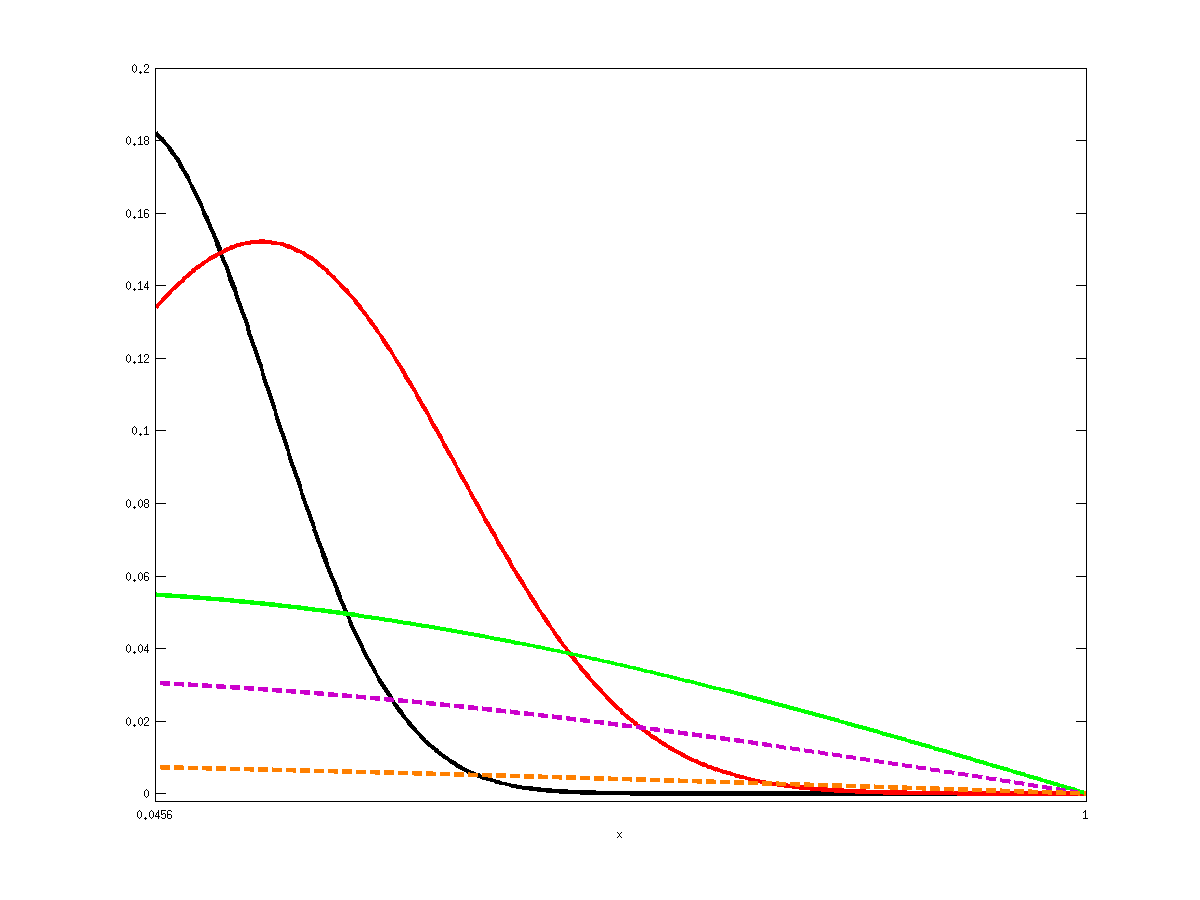}
\tabularnewline
\end{tabular}
%\par\end{centering}
%\end{turn}
\caption{Profiles of concentration in all layers at five times (continuous lines during current administration, dashed lines after suspension), for three differences of potential.
The effect of iontophoresis is to enhance penetration in the stratum corneum, and for higher voltage, even in the deeper layers (simulations based on  a uniform mesh size $h_0=h_1=h_2=h_3= 2 \cdot 10^{-3}$, current switched off after 30min).}  
\end{figure}
\end{landscape}

\begin{figure}
\centering

\includegraphics[scale=0.4]{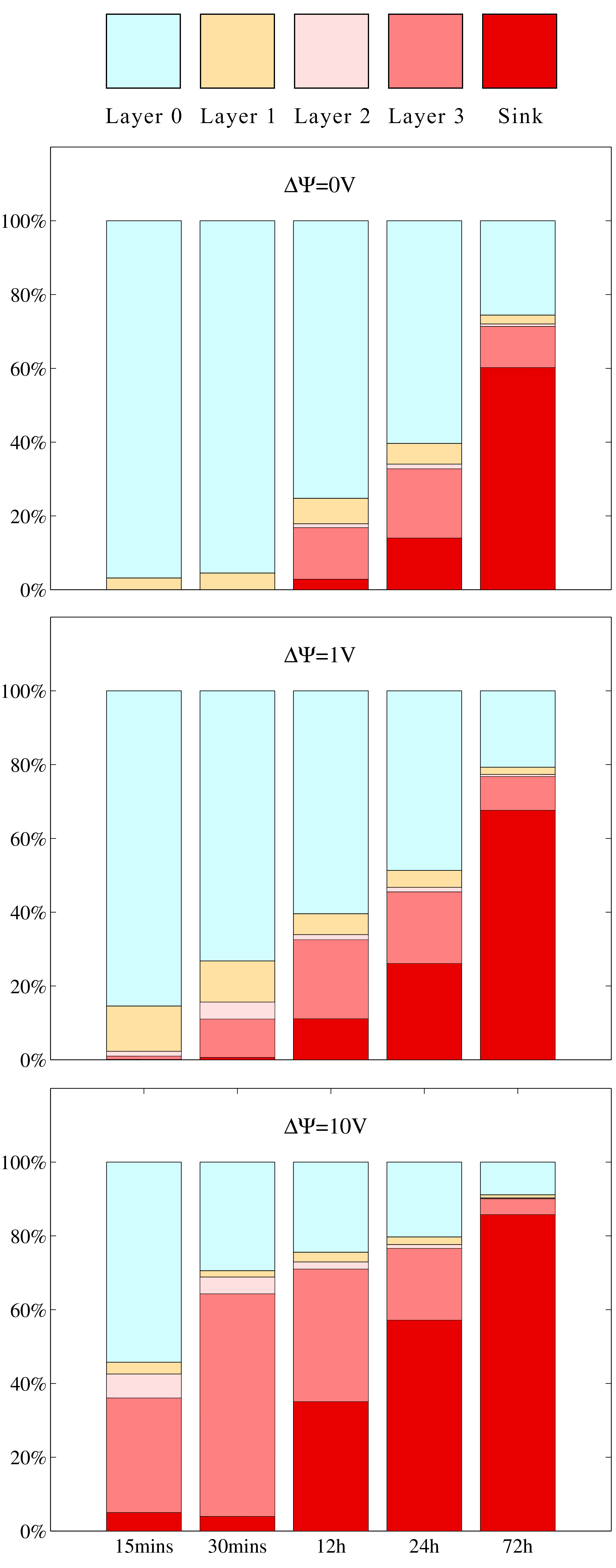}
\caption{Distribution of mass in all layers (including the vehicle (layer 0) and the sink) at five times, for three differences of potential. The heights of the rectangles
 indicate the percentage of mass retained in each layer. A  higher current 
accelerates depletion of the vehicle and
 enhances drug delivery in stratum corneum and in the deeper skin's layers. At later times most of drug results absorbed at the systemic level (current switched off after 30min). }  
\end{figure}

\end{document}